\documentclass[prd,preprint,superscriptaddress,amsmath,amssymb,nofootinbib]{revtex4}
\usepackage{graphicx}
\usepackage{dcolumn}
\usepackage{bm}
\usepackage{amssymb}
\usepackage{amsmath}
\usepackage{epsfig}    
\usepackage{color}
\usepackage{slashed}
\usepackage{hhline}
\usepackage{changes}

\def\be{\begin{equation}}
\def\ee{\end{equation}}
\newcommand{\bea}{\begin{eqnarray}}
\newcommand{\eea}{\end{eqnarray}}
\newcommand{\nn}{\nonumber}

\begin{document}

\begin{flushright}{KIAS-P21043,} {CTP-SCU/2021031}, APCTP Pre2021-015
\end{flushright}

\title{ 
Muon $g-2$, $B\to K^{(*)}\mu^+ \mu^-$ anomalies, and leptophilic dark matter 
in $U(1)_{\mu-\tau}$ gauge symmetry}

\author{ P. Ko}
\email{pko@kias.re.kr}
\affiliation{School of Physics, KIAS, Seoul 02455, Korea}
\affiliation{Quantum Universe Center, KIAS, Seoul 02455, Korea}

\author{Takaaki Nomura}
\email{nomura@scu.edu.cn}
\affiliation{College of Physics, Sichuan University, Chengdu 610065, China}

\author{Hiroshi Okada}
\email{hiroshi.okada@apctp.org}
\affiliation{Asia Pacific Center for Theoretical Physics, Pohang 37673, Republic of Korea}
\affiliation{Department of Physics, Pohang University of Science and Technology, Pohang 37673, Republic of Korea}

\date{\today}

\begin{abstract}
We propose a new class of $U(1)_{\mu-\tau}$ gauged model that can explain recent flavor anomalies such as muon $g-2$, $b\to s\mu^+ \mu^-$, as well as include a scalar dark matter candidate while satisfying all the phenomenological constraints. For this purpose, we add new vectorlike quarks and leptons as well as two inert singlet scalars, one leptophilic and the other leptophobic. In our model $b\to s\mu^+ \mu^-$ anomalies can be explained by loop induced interactions  among the SM and exotic quarks and $Z'$ gauge boson. In our numerical analysis, we show allowed region of our model to be narrow,  and it would be tested soon, for example by searching for 4 muons and dimuon + missing transverse momentum from $pp \to \mu^+ \mu^- Z'$ and $pp \to \nu_{\mu,\tau} \bar \nu_{\mu,\tau} Z'$ followed by $Z' \to \mu^+ \mu^- , \nu_{\mu,\tau} \bar \nu_{\mu,\tau} $  at the LHC. 
\end{abstract}
\maketitle

\section{Introduction}
Even after discovery of the standard model (SM) Higgs boson, there are a number of 
reasons for physics beyond the SM (BSM), such as Baryon Asymmetry of Universe (BAU), neutrino oscillations~\cite{Esteban:2020cvm}, nonbaryonic
dark matter~\cite{Planck:2015fie}, all of which are well established. 
Also there are some anomalies in the flavor sector: the ratio between rare $B$  meson 
decays of  $B^+\to K^+\mu^+ \mu^-$ and $B^+\to K^+ e\bar e$  ($R_K$)~\cite{Hiller:2003js, Bobeth:2007dw, Aaij:2014ora,Aaij:2019wad}, and muon anomalous magnetic moment (muon $g-2$)~\cite{Bennett:2006fi,Abi:2021gix}. These anomalies may call for explanations within some
new physics beyond the SM (BSM),   although some anomalies may (partly) originate from 
poorly understood SM predictions involving nonperturbative QCD.  

Especially, muon $g-2$ is known as longstanding anomaly that was initially discovered 
by E821 experiment at Brookhaven National Lab (BNL) two decades ago~\cite{Bennett:2006fi}, and has recently been confirmed 
by E989 Run 1 experiment at Fermilab(FNAL)~\cite{Abi:2021gix};
combining these measurements we obtain
\begin{align}
\Delta a_\mu =(25.1\pm5.9)\times 10^{-10},
\end{align}
whose deviation from SM prediction~\cite{Aoyama:2012wk,Aoyama:2019ryr,Czarnecki:2002nt,Gnendiger:2013pva,Davier:2017zfy,Keshavarzi:2018mgv,Colangelo:2018mtw,Hoferichter:2019mqg,Davier:2019can,Keshavarzi:2019abf,Kurz:2014wya,Melnikov:2003xd,Masjuan:2017tvw,Colangelo:2017fiz,Hoferichter:2018kwz,Gerardin:2019vio,Bijnens:2019ghy,Colangelo:2019uex,Blum:2019ugy,Colangelo:2014qya,Hagiwara:2011af}
 is 4.2 $\sigma$.

Also $R_K$ anomaly has been recently updated by the LHCb collaboration~\cite{Aaij:2021vac}: 
\begin{align}
R_K=0.846^{+0.042+0.013}_{-0.039-0.012} \quad(1.1 {\rm GeV^2}< q^2<6{\rm GeV^2}),
\end{align}
where the first (second) uncertainty is statical (systematic) one and $q^2$ is 
the dilepton invariant mass squared. 
Its deviation from the SM prediction is about 3.1 $\sigma$.
There are also discrepancies in the measurements of the angular observable $P'_5$ 
in the $B$ meson decay ($B \to K^* \mu^+ \mu^-$)~\cite{DescotesGenon:2012zf, Aaij:2015oid, Aaij:2013qta,Abdesselam:2016llu, Wehle:2016yoi}, as well as in  
$R_{K^*} = BR (B \to K^* \mu^+ \mu^-)/ BR (B \to K^* e^+ e^-)$~\cite{Aaij:2017vbb}; 
we call these as $b \to s \mu^+  \mu^-$ anomalies.

$U(1)_{\mu-\tau}$ gauge symmetry is anomaly free even without 
additional chiral fermions beyond the SM fermions~\cite{He:1990pn, He:1991qd}.  
It has been frequently applied to models that explains some of these flavor anomalies. 
Due to feature of flavor dependent symmetry, we often obtain flavor specific signals that can be 
more testable through various experiments than the SM.
Moreover, since a new gauge boson does not directly interact with electrons up to small 
knetic mixing, we can evade stringent constraints on the gauge coupling and 
gauge boson mass from LEP (or ILC) easier than the other gauged models.~\footnote{Notice here that we have to consider a constraint of neutrino trident production in gauged $U(1)_{\mu-\tau}$ models~\cite{Altmannshofer:2014pba}, and we will discuss it in the main text. }  Also the muon $g-2$ within the $U(1)_{\mu -\tau}$ gauge 
models was studied right after the BNL announcement ~\cite{Baek:2001kca}.   And the PAMELA 
$e^+$ excess was studied in the $U(1)_{\mu -\tau}$-charged dark matter with mass 
$\sim O(1)$ TeV in Ref.~\cite{Baek:2008nz}.

In fact, there is a wide variety of recent applications of $U(1)_{\mu -\tau}$ models, which 
are distinguished by additional particle contents to address flavor anomalies; Refs~\cite{Baek:2015mna, Nomura:2018cle, Nomura:2018vfz, Asai:2018ocx, Asai:2017ryy, Lee:2017ekw} for neutrino oscillations to control lepton flavor structure,  Refs~\cite{Davier:2010nc, Davier:2017zfy, Davier:2019can, Borah:2021mri, Qi:2021rhh, Singirala:2021gok, Buras:2021btx, Zhou:2021vnf, Borah:2021jzu, Chen:2021vzk, Zu:2021odn, Huang:2021nkl, Patra:2016shz, Altmannshofer:2016oaq} for muon $g-2$, Refs~\cite{Altmannshofer:2014cfa, Crivellin:2015mga, Crivellin:2015lwa, Ko:2017yrd, Kumar:2020web,Han:2019diw,Chao:2021qxq,Baek:2019qte,Chen:2017usq,Borah:2021khc} for 
$b \to s \mu^+ \mu^-$ anomalies, and (or) DM.  Qualitative features in $U(1)_{\mu-\tau}$ models for flavor anomalies are the following.
The muon $g-2$ anomaly is explained by light $Z'$ around $\mathcal{O}(10)$ to $\mathcal{O}(100)$ MeV scale or by Yukawa interactions with exotic fermions controlled by $U(1)_{\mu - \tau}$ symmetry.
For explaining $b \to s \mu^+  \mu^-$ anomalies with $U(1)_{\mu -\tau}$,  generally we need to add new particles since the corresponding gauge boson $Z'$ does not couple to quarks.
Mixing among the SM quarks and new vector-like quarks are considered to generate flavor violating $Z'$-quark coupling in refs~\cite{Crivellin:2015mga, Altmannshofer:2014cfa,Han:2019diw} while flavor dependent charge assignment in the SM quark sector is considered in ref.~\cite{Crivellin:2015lwa}.

In ref.~\cite{Ko:2017yrd}, we proposed a gauged $U(1)_{\mu - \tau}$ model in which interactions among $Z'$ and quarks are radiatively generated via one-loop diagram with 
vector-like quarks($Q'$) and scalar dark matter(DM) candidate propagating inside loop~\footnote{See also refs.~\cite{Kumar:2020web,Chao:2021qxq,Baek:2019qte,Chen:2017usq} for similar mechanism and analysis.
Also scenarios in which vector-like fermions are mixed with the SM fermions are also found e.g. in refs.~\cite{Altmannshofer:2014cfa, Navarro:2021sfb}.}.
In this work, we focused on a loop diagram associated with 
$Z'Q'\bar Q'$ interaction. 
But there are contributions associated with $Z Q' \bar Q'$ and photon$-Q' \bar Q'$ interactions, 
since vector-like quarks interact with $Z$ and photon that should be taken into account  to check 
consistency of the scenario.
Note also that we could not explain muon $g-2$ in this scenario since $Z'$ from $U(1)_{\mu - \tau}$ should be heavier than $\mathcal{O}(10)$ GeV.  Also in Ref. ~\cite{Ko:2017yrd} we introduced 
only vector-like quarks and not vector-like leptons, and the model looks non symmetric.
Thus it is interesting to add vector-like leptons in addition to vector-like quarks in the previous 
scenario to make the model look more symmetrical, to explain muon $g-2$ and consider loop diagrams associated with $Z$ and photon interactions to check consistency for $b\rightarrow s \mu^+ \mu^-$ 
anomalies.  We will also introduce two inert complex scalars $\chi_q$ (leptophobic) 
and $\chi_l$ (leptophilic) that are charged under $U(1)_{\mu - \tau}$ in such a way that Yukawa 
couplings among the vector-like quarks (leptons),  SM quarks (leptons) and the inert scalar $\chi_q$ 
($\chi_l$) are allowed by gauge symmetries. 

In the present work, we extend the previous model~\cite{Ko:2017yrd} by introducing vector-like 
leptons to explains muon $g-2$, $R_K$, and DM in a gauged $U(1)_{\mu-\tau}$ symmetry, 
where DM is assumed to be a weak interacting massive particle (WIMP).
We then estimate one loop diagrams considering $Z'$, $Z$ and photon vertices with vector-like quarks that affects $b \to s \mu^+ \mu^-$ process to explain $R_K$ anomaly and to check consistency of the model.
We also estimate muon $g-2$ and DM relic density to search for parameters explaining them. 
In addition we take all the valid constraints such as neutral meson mixings and 
$B_{s(d)}^0 \to \mu^+ \mu^-$  into consideration and show 
the allowed region in our numerical analysis.

This letter is organized as follows.
In Sec. II, we present our model and show several constraints from $B\to K^{(*)}\ell^+\ell^-$, neutral meson mixings, muon $g-2$, and (leptophilic) 
dark matter.  In Sec.III we carry out the numerical analysis and demonstrate our allowed region.
Finally, Sec. IV is devoted to the summary of our results and conclusion.

\section{Model setup and Constraints}
\begin{table}
\begin{tabular}{|c||c|c||c|c|c||c|}\hline\hline  
& ~$Q'$~  & ~$\chi_q$~& ~$L'$~& ~$E$~  & ~$\chi_\ell$ ~&~ $\varphi$~  \\\hline\hline 
$SU(3)_C$ & $\bm{3}$  & $\bm{1}$ & $\bm{1}$ & $\bm{1}$ & $\bm{1}$ & $\bm{1}$  \\\hline 
$SU(2)_L$ & $\bm{2}$  & $\bm{1}$ & $\bm{2}$ & $\bm{1}$ & $\bm{1}$ & $\bm{1}$  \\\hline 
$U(1)_Y$   & $\frac16$ & $0$  & $-\frac12$& $-1$ & $0$ & $0$  \\\hline
$U(1)_{\mu-\tau}$   & $q_x$ & $q_x$ & $\frac{q_x}2$ & $\frac{q_x}2$&  $\frac{q_x}2-1$ & ${q_\varphi}$  \\\hline
\end{tabular}
\caption{ 
Charge assignments of the new fields $Q'$, $L'\equiv [N',E']^T$, $E$, $\chi_q$ and $\chi_\ell$  
under $SU(3)_C\times SU(2)_L\times U(1)_Y\times U(1)_{\mu-\tau}$. 
Here $Q'$, $L'$ and $E$ are vector-like fermions. $\chi_{q}$ and $\chi_{\ell}$ are complex 
inert scalar bosons either of which can be considered as a DM candidate.
{ $\varphi$ plays a role in breaking the $U(1)_{\mu-\tau}$ symmetry spontaneously 
by the nonzero VEV; $\langle \varphi \rangle = v_\varphi/\sqrt{2}$.}}
\label{tab:1}
\end{table}

\begin{figure}[t]
\begin{center}
\includegraphics[width=75mm]{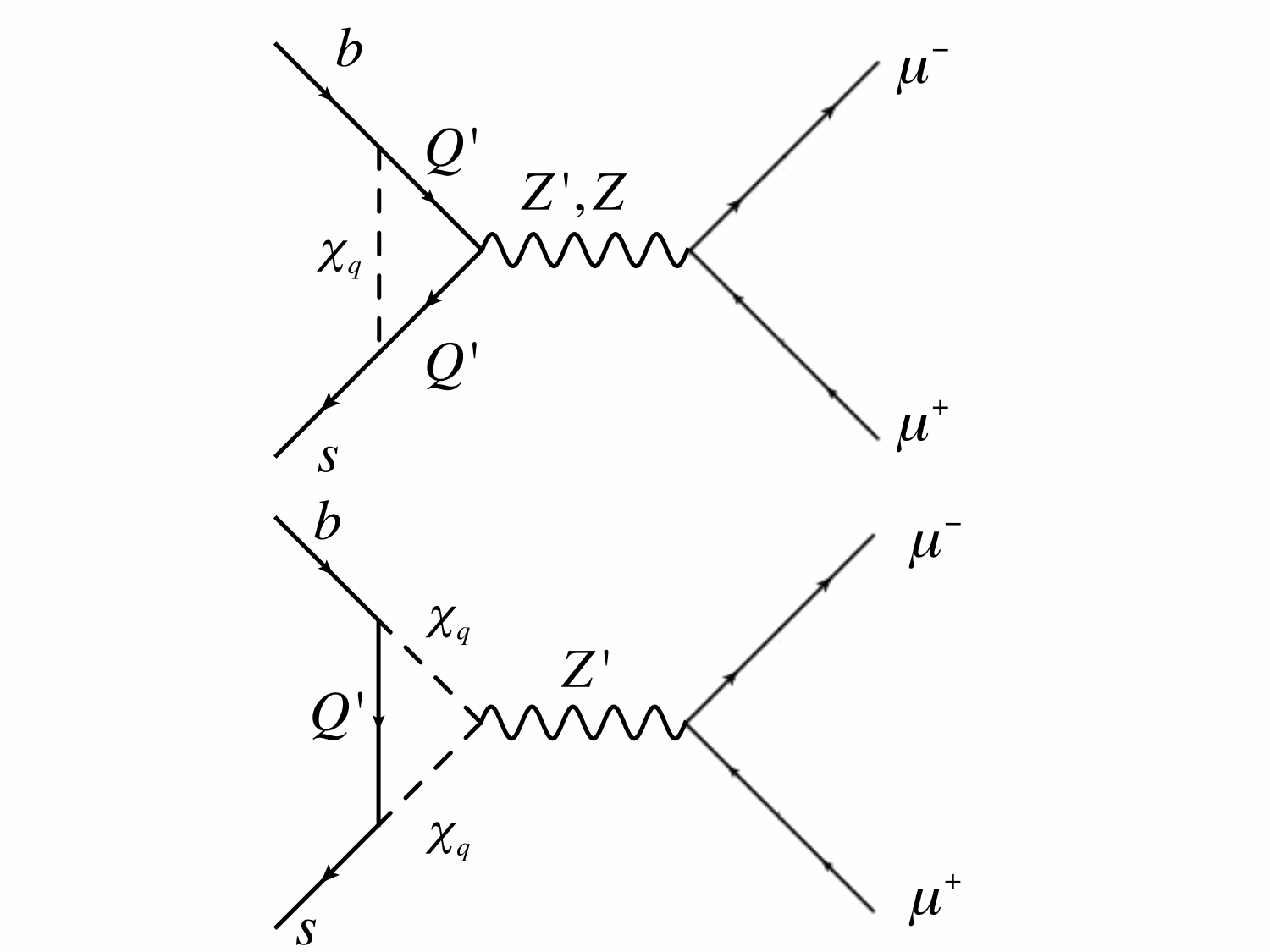} \qquad
\includegraphics[width=75mm]{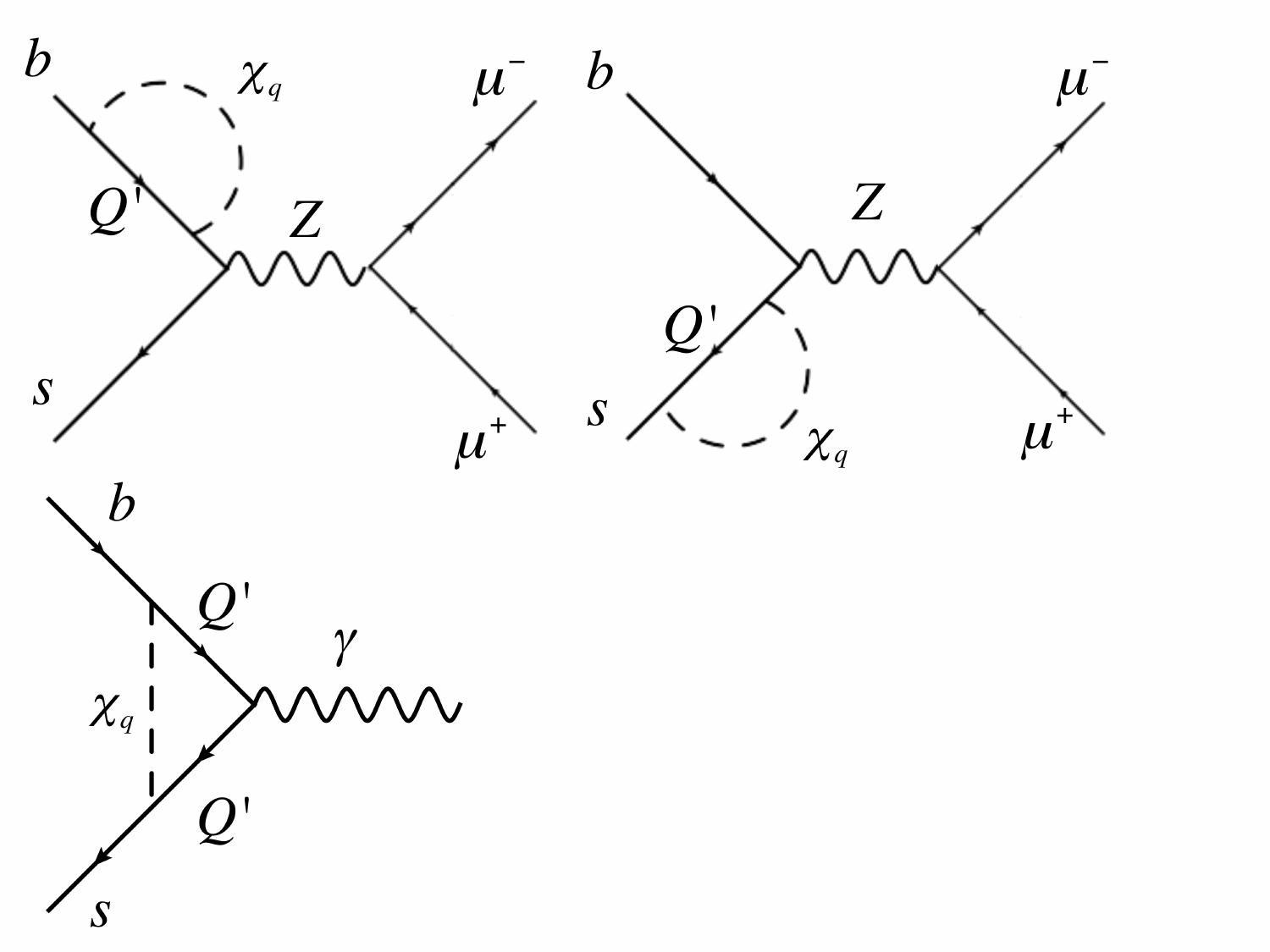}
\caption{The diagrams introducing effective coupling for $Z'_\mu (Z_\mu) \bar b \gamma^\mu P_L s + h.c.$ and $F^{\mu \nu} \bar b \sigma_{\mu \nu} P_{L,R} s + h.c.$ interaction. } 
  \label{fig:diagram}
\end{center}\end{figure}

In this section we define our model. 
We introduce a vector-like exotic quark $Q'\equiv [U',D']^T$ and lepton $L'\equiv [N',E']^T$ under isospin doublet and a singly charged-lepton $E$ under isospin singlet, two complex scalar bosons $\chi_q$ and $\chi_\ell$, and a boson $\varphi$,
where $\chi_q$($\chi_\ell$) couples only to quark(lepton) sector respectively.  { The singlet  
scalar $\varphi$ plays a role in breaking the $U(1)_{\mu-\tau}$ symmetry spontaneously 
developing the nonzero vacuum expectation value (VEV); $\langle \varphi \rangle = v_\varphi/\sqrt{2}$.
Here we take the $U(1)_{\mu-\tau}$ charge of $\varphi$ so as to retain the inert features of $\chi_{q,\ell}$;
we forbid terms $\chi_{l,q}^{(*)} (\varphi^{(*)})^n$ with $n$ being integer.
This can be achieved by imposing conditions for the charge of 
$\varphi$ as follows: 
\begin{align}
\label{eq:cond1}
q_\varphi \neq \pm \frac1n \left( \frac{q_x}{2} - 1 \right), \ \pm \frac1n q_x.  
\end{align}
In addition we forbid these operators, $\bar Q_L Q'_R \varphi^{(*)}$, $\bar L_{L_\ell} L'_L \varphi^{(*)}$ and $\bar \ell_R E_L \varphi^{(*)}$ $(\ell = e, \mu, \tau)$ and their Hermitian 
conjugates, so that newly added vector-like fermions and the SM ones do not mix.
In fact, $\bar Q_L Q'_R \varphi^{(*)}$, $\bar L_{L_{e, \mu}} L'_L \varphi^{(*)}$ and $\bar e_R(\bar \mu_R) E_L \varphi^{(*)}$ are already forbidden by the conditions in Eq.~\eqref{eq:cond1}.
To forbid remaining operators we require additional conditions, 
\begin{equation}
\label{eq:cond2}
q_\varphi \neq  \pm \left( \frac{q_x}{2} + 1 \right).
\end{equation}
}
The SM Higgs is denoted by $H$ and its VEV is given by $[0,v/_H/\sqrt2]^T$.
$Q'$ as well as $\chi_{q}$ is requested to induce sizable $B\to K^{(*)}\ell\bar\ell$ anomalies, 
$E$ and $L'$ as well as $\chi_{\ell}$ are required to get sizable lepton anomalous magnetic moment.~\footnote{With $E$ only, one cannot get sizable lepton anomalous magnetic moment, since the mass of $E$ has to be very small at about 200 GeV, which would be ruled out by LHC. $L'$ plays a role in evading chiral suppression and thus sizable values are obtained.}
Charge assignments of these new fields are summarized in Table~\ref{tab:1}.
Then, the valid Lagrangian under these symmetries is given by 
\begin{align}
-\mathcal{L}_{\rm VLF+ \chi}
=&  M_{Q'} \bar Q' Q' + M_{E} \bar E' E'  + { M_{L'} \bar L'_L L'_R} + y \overline{L'_L} H E_R+ m_{\chi_q}^2 |\chi_q|^2 + m_{\chi_\ell}^2 |\chi_\ell|^2
  \nn \\
  & +( f_{i} \overline{Q'_R} Q_{L_i}\chi_q 
+ g_{\mu} \overline{\mu_R} E_{L} \chi_\ell^* 
+ h_{\mu} \overline{L_{L_\mu}} L'_R \chi_\ell^* 
+ {\rm h.c.}),
\label{Eq:lag-flavor}
\end{align} 
where $i=1-3$ 
are generation indices, {$Q_{L}$'s are the SM quark doublets and 
$\mu_R$ is the right-handed muon.
We have abbreviated kinetic terms for simplicity.

The scalar potential in our model is given by
\begin{align}
V =  -M_H^2 H^\dagger H - M^2_\varphi \varphi^* \varphi + \lambda_H (H^\dagger H)^2 + \lambda_\varphi (\varphi^* \varphi)^2 + \lambda_{H \varphi} (H^\dagger H)(\varphi^* \varphi),
\end{align}
where we have abbreviated scalar potential associated with $\chi_{q,\ell}$ for simplicity.
We expand scalar fields around its VEVs as
\begin{align}
H = 
\begin{pmatrix}
w^+ \\
\frac{1}{\sqrt{2}} (v + \tilde{h} + i G_Z ) 
\end{pmatrix}, \quad
\varphi &= \frac{1}{\sqrt{2}} (v_\varphi + \tilde{\phi} + i G_{Z'}),
\label{eq:scalar-fields}
\end{align}
where $\tilde{h}$ and $\tilde{\phi}$ correspond to CP-even physical scalar boson states while $w^+$, 
$G_Z$ and $G_{Z'}$ are the Nambu-Goldstone(NG) bosons which are absorbed by the weak gauge bosons $Z,~W$ and $Z'$.
The mass matrix for physical scalar bosons is obtained as 
\begin{align}
M^2 = 
\begin{pmatrix}
\lambda_H v^2 & \lambda_{H\varphi} v v_\varphi \\
\lambda_{H\varphi} v v_\varphi & \lambda_\varphi v_\varphi^2
\end{pmatrix}.
\end{align}
The matrix is diagonalized by acting an orthogonal matrix $U$ such that
\begin{align}
 U^T M^2 U = \mathrm{diag}(m_h^2, m_H^2),
\end{align}
where 
\begin{align}
U = 
\begin{pmatrix}
\cos\alpha & \sin\alpha \\
-\sin\alpha & \cos\alpha
\end{pmatrix},
\end{align}
and the scalar mixing angle $\alpha$ is given by
\begin{align}
\tan 2\alpha = \frac{2 \lambda_{H \varphi} v v_\varphi}{\lambda_H v^2 - \lambda_\varphi v_\varphi^2}. \label{eq:scalar-mixing}
\end{align}
Then mass eigenstates are given by
\begin{align}
\begin{pmatrix}
h \\
\phi
\end{pmatrix}
=
U^T
\begin{pmatrix}
\tilde{h} \\
\tilde{\phi}
\end{pmatrix}
=
\begin{pmatrix}
\cos\alpha & -\sin\alpha \\
\sin\alpha & \cos\alpha
\end{pmatrix}
\begin{pmatrix}
\tilde{h} \\
\tilde{\phi}
\end{pmatrix},
\label{eq:scalar-eigenstates}
\end{align}
where $h$ corresponds to the SM-like Higgs boson.

Exotic singly-charged fermions $(E,E')$ mix each other after spontaneous electroweak symmetry breaking, and its mass matrix is given by
\begin{align}
{\cal M}_E=
\left(\begin{array}{cc}
M_E & m_E \\
m_E &M_{L'}\\
\end{array}\right), 
\end{align}
where  $m_E\equiv y v /\sqrt2$.
It is diagonalized by a unitary matrix $V_E$ as ${\rm diag}(M_1,M_2)=V_E^\dag {\cal M}_E V_E$,
 and $V_E$ is given by
\begin{align}
V_E=
\left(\begin{array}{cc}
c_c & -s_c \\
s_c &c_c \\
\end{array}\right),  \quad \tan 2 \theta_c = \frac{2 m_E}{M_{E} - M_{L'}},
\label{eq:mixing}
\end{align}
where $c_c(s_c)$ is shorthand symbol of $\cos\theta_c(\sin\theta_c)$.
We write mass eigenstates as $E_{1}$ and $E_2$ whose masses are $M_{1}$ and $M_2$.

\section{Phenomenological Constraints} 
\subsection{Wilson coefficients for $B \to K^{(*)} \ell^+ \ell^-$ decay}

Here we derive Wilson coefficients associated with $B \to K^{(*)} \ell^+ \ell^-$ processes derived from diagrams in Fig.~\ref{fig:diagram}.
Anomalies observed in $B \to K^{(*)} \ell^+ \ell^-$ decays can be explained by the shift of the Wilson coefficient $C_9$ which corresponds to the operator $(\bar s\gamma_\mu P_L b)(\mu\gamma^\mu \mu)$.
In our model, the effective coupling for $Z'_\mu \bar b \gamma^\mu P_L s + h.c.$ is induced 
at one loop level as shown in Fig.~\ref{fig:diagram} with the Yukawa coupling in Eq.~(\ref{Eq:lag-flavor}).
Then the effective Hamiltonian $(\bar s\gamma_\mu P_L b)(\bar{\mu}\gamma^\mu \mu)$ arises from $Z'$ mediation and the contribution to Wilson coefficient $ \Delta C_9^{\mu \mu}(Z')$ is obtained as:
\begin{align}
& \Delta C_9^{\mu \mu}(Z')  \simeq -\frac{q_x g'^2}{4 (4\pi)^2 m_{Z'}^2 C_{\rm SM}}
f^*_{3}f_{2} ,
\nn \\
& { C_{\rm SM}  \equiv \frac{V_{tb}V^*_{ts} G_{\rm F}\alpha_{\rm em}}{\sqrt2 \pi}, } 
\label{eq:c9}
\end{align}
where 
$V_{tb}\approx0.999$, $V_{ts}\approx-0.040$ are  the 3-3 and 3-2 elements of CKM matrix respectively,
$G_{\rm F}\approx 1.17\times 10^{-5}$ GeV is the Fermi constant, $\alpha_{\rm em}\approx1/137$ is the 
electromagnetic fine-structure constant, 
{$m_b\approx 4.18$ GeV and $m_s\approx {0.095}$ GeV are respectively the bottom and strange quark masses 
given in the $\overline{MS}$ scheme at a renormalization scale $\mu = 2$ GeV~\cite{pdg}.}
Notice here that we have ignored $m_b$ and $m_s$ in the formula for    
$C_9$ in Eq.~(\ref{eq:c9}), since they are much smaller than $M_{Q'}$ and $m_{\chi_q}$ 
in our scenario.  
The $ \Delta C_{9}^{\mu \mu}$ allowed within 2$\sigma$ range is given by~\cite{Alguero:2021anc}:
\begin{align}
0.83\le -\Delta C_{9}^{\mu \mu}\le 1.52.
\end{align}

We also obtain lepton universal Wilson coefficient $\Delta C_9(Z)$ from the diagram where $Z'$ is replaced to $Z$.
Another lepton universal Wilson coefficient $\Delta C_{10}(Z)$ associated with the corresponding operator $(\bar s\gamma_\mu P_L b)(\ell \gamma^\mu\gamma_5 \ell)$
is also arisen from $Z$ mediation. The formulas of them are given by
\begin{align}
& \Delta C_9^{}(Z) \simeq 
\frac{g_2^2 f^*_{3}f_{2} }{4 (4\pi)^2 m_{Z}^2 c_w^2 C_{\rm SM}}
\left(-\frac12 + 2 s^2_w \right)
\left(-\frac12 + \frac13 s^2_w \right), \\
& \Delta C_{10}^{}(Z) \simeq 
\frac{g_2^2 f^*_{3}f_{2} }{8 (4\pi)^2 m_{Z}^2 c_w^2 C_{\rm SM}}
\left(-\frac12 + \frac13 s^2_w \right).
\end{align}
The $ \Delta C_{10}^{}$ contributes to $B_s^0 \to \mu^+ \mu^-$ process.
Here we consider measurement of $B_s^0 \to \mu^+ \mu^-$ branching ratio~\cite{LHCb:2021vsc,LHCb:2021awg}
\begin{equation}
BR(B_s^0 \to \mu^+ \mu^-)^{\rm exp} = (3.09^{+0.46 + 0.15}_{-0.43-0.11}) \times 10^{-9},
\end{equation}
where the first and second uncertainties are statistical and systematic respectively.
We compare the experimental value with the theoretical one estimated by~\cite{Hiller:2014yaa}
\begin{equation}
BR(B_s^0 \to \mu^+ \mu^-)^{\rm th} = |1-0.24 \Delta C_{10}^{\mu \mu}|^2 BR(B_s^0 \to \mu^+ \mu^-)^{\rm SM},
\end{equation} 
where $BR(B_s^0 \to \mu^+ \mu^-)^{\rm SM} = (3.65 \pm 0.23) \times 10^{-9}$ is the SM prediction~\cite{Bobeth:2013uxa}.
$\Delta C_{10}^{\mu \mu}$ is restricted by the data of the process and $B \to K^{(*)} \ell^+ \ell^-$.
For example, allowed range at 2$\sigma$ is given by~\cite{Alguero:2021anc} including data of 
$B_s^0 \to \mu^+ \mu^-$ decay:  $-0.11\le \Delta C_{10}^{\mu \mu}\le 0.49$.
 In fact this range will be modified as we have lepton flavor universal $\Delta C_9 (Z)$ and $\Delta C_{10}(Z)$ effects.
 Here we impose $0 < \Delta C_{10}(Z) < 0.79$ to satisfy measurement of 
 $BR(B_s^0 \to \mu^+ \mu^-)$;  we assume $\Delta C_{10}(Z) > 0$ since it is preferred by
 data.

Furthermore the Wilson coefficient $C_7$ is modified by $\Delta C_{7}$:
\begin{align}
 \Delta C_7 = \frac{f_2 f^*_3}{2} \int [dX]^3 \frac{xy}{x(x-1)m_b^2 + x m^2_{\chi_q} +(y+z) M^2_{Q'}} \left( \frac{4 G_F }{\sqrt{2}} V_{tb} V^*_{ts} \right)^{-1}, 
\end{align}
where $\int [dX]^3 \equiv \int_0^1 dx dy dz \delta(1-x-y-z)$.
In our numerical analysis we apply the constraints~\cite{Alguero:2021anc}:
\begin{align}
-0.01 \le \Delta C_{7} \le 0.03.
\end{align}
Note that Wilson coefficient $\Delta C_{7'}$ is also induced but it is suppressed by $m_s/m_b$ factor compared to $C_7$.

\subsection{$M-\overline M$ mixing}
      
The exotic vector-like quarks and the complex scalar DM $\chi$ induce 
the neutral meson ($M$)-antimeson ($\overline M$) mixings such as $K^0-\bar K^0$, $B_d-\bar B_d$, 
$B_s-\bar B_s$, and  $D^0-\bar D^0$ from the box type one-loop diagrams.  
The formulae for the mass splitting are  respectively given by~\cite{Gabbiani:1996hi}
\begin{align}
 \Delta m_K & \approx
\sum_{a,b=1}^3
 |f_1 f_2|^2 G^K_{box}[m_{\chi_q}, M_{Q'}]   \lesssim 3.48\times10^{-15} \ [{\rm GeV}],
\label{eq:kk}\\
 \Delta m_{B_d} & \approx
\sum_{a,b=1}^3 |f_1 f_3|^2 
 G^{B_d}_{box}[m_{\chi_q}, M_{Q'}]  \lesssim 3.36\times10^{-13} \ [{\rm GeV}],\\
 \Delta m_{B_s}  & \approx
\sum_{a,b=1}^3 |f_2 f_3|^2 
G^{B_s}_{box}[m_{\chi_q}, M_{Q'}]  \lesssim 1.17\times10^{-11} \ [{\rm GeV}],\\
 \Delta m_D & \approx
\sum_{a,b=1}^3|f_1 f_2|^2 
G^{D}_{box}[m_{\chi_q}, M_{Q'}]   \lesssim 6.25\times10^{-15} \ [{\rm GeV}],\label{eq:dd} \\
 G^M_{box}(m_1,m_2)  & =
\frac{m_M f_M^2}{3(4\pi)^2}
 \int [dX]^3 \frac{x  }{x m^2_1+(y+z) m_2^2 } \nn \\
& = 
\frac{m_M f_M^2}{6(4\pi)^2}
\left[\frac{m_1^4 - m_2^4+4 m_1^2 m_2^2 \ln\left[\frac{m_2}{m_1}\right]}{(m_1^2-m_2^2)^3}\right]
 \hspace{4cm}(m_1\neq m_2)
,\label{eq:mmbar-mix}
\end{align}
where relevant quarks $(q,q')$ are respectively $(d,s)$ for $K$,  $(b,d)$ for ${B_d}$,  $(b,s)$ for ${B_s}$, and  
$(u,c)$ for $D$, each of the last inequalities of the above equations 
represent the upper bound from the experimental values \cite{pdg}, and
$f_K\approx0.156$ GeV, $f_{B_d(B_s)}\approx0.191(0.200)$ GeV, $f_{D}\approx0.212$ GeV,
 $m_K\approx0.498$ GeV,  $m_{B_d(B_s)}\approx5.280(5.367)$ GeV, and  $m_{D}\approx 1.865$ GeV.

\subsection{ Muon anomalous magnetic dipole moment}
Recently,  E989 Collaboration at Fermilab reported the new result on the muon $(g-2)$ 
\cite{Abi:2021gix}: 
\begin{align}
a^{\rm FNAL}_\mu =116592040(54) \times 10^{-11},
\label{exp_dmu}
\end{align}
Combining it with the previous BNL result, the new result 
on the muon $(g-2)$ shows a deviation from the SM prediction at 4.2 $\sigma$ level:
\begin{align}
\Delta a^{\rm new}_\mu = (25.1\pm 5.9)\times 10^{-10} ,
\label{exp_dmu}
\end{align}
which may be a herald for new physics beyond the SM.

In our model, the new dominant contribution arises from the loop diagrams 
involving the vector-like leptons and leptophilic DM $\chi_l$.  The dominant terms involve  
a product of $g_\mu$ and $h_\mu$~\footnote{The contributions proportional to $|g_\mu|^2$ or $|h_\mu|^2$ cannot be dominant because of the chiral suppression. Thus, we neglect those terms for simplicity.}, 
and its form is given by
\begin{align}
\Delta a^{\rm new}_\mu &=\frac{m_\mu}{(4\pi)^2}s_c c_c g_\mu h_\mu
\left[ \frac{F(q_1, r_1)}{M_1} -  \frac{F(q_2, r_2)}{M_2} \right] ,\\
F(q,r)&\equiv \int_0^1 dx\int_0^{1-x} dy
\frac{1-2y}{1-x + q^2 (x^2-x) + r^2 x  }
\approx
\frac{-1+r^4-2r^2\ln(r^2)}{2(-1+r^2)^3} +{\cal O}(q^2),
\label{exp_dmu}
\end{align}
where $q_i\equiv \frac{m_\mu}{M_i}(<<1)$ and $r_i\equiv \frac{m_{\chi_\ell}}{M_i}$. 
And $M_i(i=1,2)$ and $s_c (c_c)$ are the mass eigenstates for the exotic singly-charged fermions 
and their mixings given in Eq.~\eqref{eq:mixing}, respectively.

\subsection{Oblique parameters}
The vector-like fermions and scalars contribute to oblique parameters through vacuum polarization diagram for electroweak gauge bosons.
Here we consider $S$- and $T$-parameters which would constrain our parameter space.
They are given by~\cite{Peskin:1990zt}
\begin{align}
& \alpha_{em} S = 4 e^2 \left( \sum_{\rm VLF} \left[ \frac{d}{d q^2} \Pi_{33}^{\rm VLF} - \frac{d}{d q^2} \Pi_{3Q}^{\rm VLF} \right]_{q^2 = 0} + \left[ \frac{d}{d q^2}  \Pi_{33}^{h,H} -  \frac{d}{d q^2}  \Pi_{3Q}^{h,H} \right]_{q^2 = 0} \right), \\ 
& \alpha_{em} T = \frac{e^2}{s_W^2 c_W^2 m_Z^2} \left( \sum_{\rm VLF} \left[ \Pi^{\rm VLF}_{\pm}(q^2) - \Pi^{\rm VLF}_{33}(q^2) \right]_{q^2 =0} 
+ \left[ \Pi^{h,H}_{\pm}(q^2) -  \Pi^{h,H}_{33}(q^2) \right]_{q^2 =0} \right),
\end{align}
where superscript VLF distinguishes contributions from different combination of our vector-like fermions $\{E_1, E_2, N, U, D\}$, and $\Pi_{33,3Q,\pm}$ are obtained from 
vacuum polarization diagrams which are summarized in the Appendix; here $\Pi^{h,H}(33,3Q,\pm)$ indicates beyond the standard model contribution.
We impose the constraint on new physics contributions to the $S$- and $T$-parameters given by~\cite{pdg}
\begin{align}
\Delta S \le  0.00 \pm 0.07 \\
\Delta T \le 0.05 \pm 0.06
\end{align} 
where we fix $U$-parameter to be zero.

\subsection{ Dark matter}  
In our scenario, $U(1)_{\mu - \tau}$ is a good global symmetry of the model Lagrangian at 
renormalizable level, and the lightest $U(1)_{\mu - \tau}$ particle with spin-0 would make 
a good DM candidate in addition to the $\nu_\mu$ and $\nu_\tau$. 
Then the lighter field of two complex scalars $\chi_q$ (hadro-philic or leptophobic)
and $\chi_\ell$ (leptophilic) can be DM candidate. 

 In fact DM physics in our model is very rich. 
For generic choices of $q_x$ and $q_\varphi$, $\chi_q$ and $\chi_l$ would be separately stable, 
and we are with two-component DM models.   However this would be no longer true for special 
choices of $q_\varphi$ and $q_x$. For example, the following operators would be gauge invariant 
while satisfying Eqs.~\eqref{eq:cond1} and \eqref{eq:cond2}:
\begin{align}
& \chi_l \chi_q \varphi , ~{\rm if}~ q_\varphi = 1 + \frac{3q_x}{2} ,
\\
& \chi_l \chi_q^\dagger \varphi^2 , ~{\rm if}~ q_\varphi = \frac{1}{2} \left( 1 + \frac{q_x}{2} \right) .
\end{align}
Then $\chi_q$ and $\chi_l$ will mix with each other, and we would end up with a single 
component DM model. In fact the mixing can occur without $\varphi$: $\chi_l \chi_q$ is allowed 
if $q_x = 2/3$. 
Another interesting possibility is an operator $\chi_l^{\dagger 2} \chi_q \varphi^\dagger$ that is 
allowed for $q_\varphi = 2$.  Then after $U(1)_{\mu - \tau}$ symmetry breaking, the effective 
$\chi_l^{\dagger 2} \chi_q$ operator is induced, so that leptophobic DM $\chi_q$ will decay into 
a pair of leptophilic DM $\chi_l$.

In this paper, we shall ignore all these interesting possibilities, and
focus on the leptophilic $\chi_\ell$ as a DM candidate that is simply denoted by $\chi$
~\footnote{See Ref.~\cite{Ko:2017yrd} for the case where $\chi_q$ is DM.}.
Then, it annihilates into the SM leptons via $\chi \chi^* \to\ell^+ \ell^-$ with $\psi_{1,2}^+$ propagator, 
and $\chi \chi^* \to Z' \to \{ \mu^+ \mu^-, \tau^+ \tau^-, \nu_{\mu, \tau} \bar{\nu}_{\mu, \tau} \}$,~\footnote{Here, the Yukawa contribution gives an $s$-wave dominant annihilation cross section, while the $Z'$ mediated cross section is  the $p$-wave dominant.} indicating our DM is leptophilic.
In fact there are annihilation mode of $\chi \chi^* \to Z' Z'$ if kinematically allowed.
In our analysis we do not consider this mode assuming $m_\chi < m_{Z'}$ to simplify our analysis.
The relic density of DM is given by
\begin{align}
&\Omega h^2
\approx 
\frac{1.07\times10^9}{\sqrt{g_*(x_f)}M_{Pl} J(x_f)[{\rm GeV}]},
\label{eq:relic-deff1}
\end{align}
where $g^*(x_f\approx25)\approx100$, $M_{Pl}\approx 1.22\times 10^{19}$,
and $J(x_f) (\equiv \int_{x_f}^\infty dx \frac{\langle \sigma v_{\rm rel}\rangle}{x^2})$ is found as
\begin{align}
&
J(x_f)=\int_{x_f}^\infty dx\left[ \frac{\int_{4m_\chi^2}^\infty ds\sqrt{s-4 m_\chi^2} (\sigma v_{\rm rel}) K_1\left(\frac{\sqrt{s}}{m_\chi} x\right)}{16  m_\chi^5 x [K_2(x)]^2}\right], \nn \\
& (\sigma v_{\rm rel}) =  (\sigma v_{\rm rel})_{\mu\bar\mu} + (\sigma v_{\rm rel})_{\tau\bar\tau} + (\sigma v_{\rm rel})_{\nu_\mu \bar{\nu}_\mu +\nu_\tau \bar{\nu}_\tau} ,\\
\label{eq:sigma1}
&(\sigma v_{\rm rel})_{\mu\bar\mu} \simeq \frac{g'^4 q_x^2 s (s-4m_\chi^2)}{24\pi (s-m_{Z'}^2)^2}
+
\frac{(s_c c_c h_\mu g_\mu)^2}{16\pi} \int_0^\pi d\theta \sin\theta 
\left(\frac{M_1}{t-M_1^2} - \frac{M_2}{t-M_2^2}\right)^2 s,\\
&(\sigma v_{\rm rel})_{\tau\bar\tau}= \frac{g'^4 q_x^2  (s+2 m_\tau^2) (s-4m_\chi^2)}{24\pi (s-m_{Z'}^2)^2} \sqrt{1-\frac{4m_\tau^2}{s}}, \\
&  (\sigma v_{\rm rel})_{\nu_\mu \bar{\nu}_\mu +\nu_\tau \bar{\nu}_\tau} =  \frac{g'^4 q_x^2 s (s-4m_\chi^2)}{24\pi (s-m_{Z'}^2)^2}.
\label{eq:relic-deff2}
\end{align}
Here  $s$ and $t$ are  Mandelstam variables, and $K_{1,2}$ are the modified  Bessel functions of the second kind of order 1 and 2, respectively.
We write second term of Eq.~\eqref{eq:sigma1} in terms of integration by angle $\theta$ since analytic form after integration is complicated as $t$ depends on $\theta$.
Note that integration by $\theta$ is carried out since corresponding amplitudes only depend on $s$.
In addition we ignore interference term between $Z'$  and $\Psi^\pm_{1,2}$ mediated diagrams since it is suppressed by $m_\mu/m_\chi$.
In our numerical analysis below, we use the current experimental 
range for the relic density: $0.11\le \Omega h^2\le 0.13$~\cite{Ade:2013zuv}.

{\it Direct detection}: We have a spin independent scattering cross section that is 
constrained by several experiments such as XENON1T~\cite{XENON:2015gkh}. 
The dominant contribution is Higgs portal via  interactions in the scalar potential.
However we simply assume the Higgs portal coupling $c_{\chi_\ell\chi_\ell h}$ for $\chi_l-\chi_l-h_{\rm SM}$ to be tiny 
enough; $c_{\chi_\ell\chi_\ell h} \lesssim 10^{-3}$~\cite{Arcadi:2019lka},  in order 
to evade the direct detection via Higgs exchanges.

\section{Numerical analysis \label{sec:numerical}}
In this section, we perform the numerical analysis.
Relevant free parameters are scanned to search for parameter points satisfying phenomenological requirements discussed in the previous section.
The ranges of the other input parameters are set to be as follows: 
\begin{align}
& g' \in [0.001, 0.1], \ |q_x| \in[0.1,\sqrt{4\pi}/g'],\ 
\{f_{1,2,3}, g_{e,\mu}, h_{e,\mu}\}  \in[10^{-2},1],\  m_{Z'}\in [m_{\chi_l}, 150]\ [{\rm GeV}],\  \nn \\
&m_{\chi_\ell} \in [1,100]\ [{\rm GeV}], \ m_{\chi_q} \in [1.2 m_{\chi_\ell},10^4]\ [{\rm GeV}], \ \{M_{E,L'}\}\in [10^2,10^4]\ [{\rm GeV}], \    \\
& m_{E} \in [10,100]\ [{\rm GeV}], \ M_{Q'} \in [10^3,10^4]\ [{\rm GeV}] , \nn
\end{align}
where we take perturbation limit $|q_xg'|\le\sqrt{4\pi}$, and choose $M_1 < M_2$, and take $m_{\chi_\ell} < 1.2 \times (M_1, m_{\chi_q})$ for simplicity so that we can 
evade contributions from co-annihilation processes. Here we require 1000 GeV$\le M_{Q'}$ to avoid constraints from direct search of vector-like quarks at the LHC.
 For vector-like lepton, constraint $m_{VLL} \gtrsim 700$ GeV is given for vector-like lepton decaying into the SM lepton with $W/Z$ boson~\cite{Sirunyan:2019ofn}. 
But here we just consider LEP limit of $m_{VLL} \gtrsim100$ GeV~\cite{Achard:2001qw} since our vector-like lepton decays into lepton and $\chi_l$. 
In addition we impose the constraint of $550$ GeV $\lesssim m_{Z'}/g'$ which arises from the neutrino trident production~\cite{Altmannshofer:2014pba}.
{ Here we take scalar mixing $\sin \alpha$ to be sufficiently small so that scalar boson contribution to oblique parameters can avoid constraints since it is proportional to $\sin^2 \alpha$.
{ We find that new scalar contribution to $S$ and $T$ parameter is sufficiently small for $\sin \alpha < 0.1$.}
Moreover, taking $m_E < M_{E,L'}$, we find that constraint from $S$- and $T$-parameters can be easily avoided where $S$ and $T$ are found to be less than $\sim 0.05$ and $0.01$, respectively.}

Firstly, in Fig.~\ref{fig:C9C10}, we show $C_{9,10}$ values for parameter sets satisfying 
the constraints of neutral meson-antimeson mixings $\Delta m_K,\Delta m_{B_d},\Delta m_{B_s},
\Delta m_D$, correct relic density $0.11\le \Omega h^2\le 0.13$ and $0< \Delta C_{10}(Z) < 0.79$; 
here we show region with $\Delta  C_{9}^{\mu \mu}(Z') < 0$ and $|\Delta C_{9,10}(Z)|$ where region 
with $\Delta  C_{9}^{\mu \mu}(Z') > 0$ and $\Delta C_{9,10} >(<) 0$ similarly appear. 
We find that $|\Delta  C_9(Z)|$ tends to be small due to suppression factor of $(2 s^2_w - 1/2)$.
Then the value of $\Delta C_9^{\mu \mu}$ can be $0.83\le -\Delta C_9^{\mu\mu}\le 1.52$ indicated 
by region between dashed line which is obtained from global analysis in Ref.~\cite{Alguero:2021anc} 
(see also Ref.~\cite{Carvunis:2021jga}).
Note that non-zero $\Delta C_{10}$ helps to fit $B_s \to \mu \mu$ data.
In fact non-zero $\Delta C^{ee}_{9,10}(Z)$ affect the region from global fit but we still use the region
as our reference.    We also find that $C_7$ is tiny in our allowed region as $|C_7|< 0.001$ and 
the  corresponding plot is omitted.

In Fig.~\ref{fig:amu}, we show our values of $\{M_1, a_\mu \}$ for the parameter sets the same as Fig.~\ref{fig:C9C10}.
We can obtain  $\Delta a^{\rm new}_\mu = (25.1\pm 5.9)\times 10^{-10}$ at 1(2)$\sigma$ for $M_1 \lesssim 2000(2700)$ GeV.
Here, the dashed line denotes the allowed region within 2$\sigma$.

\begin{figure}[t]
\begin{center}
\includegraphics[width=60.0mm]{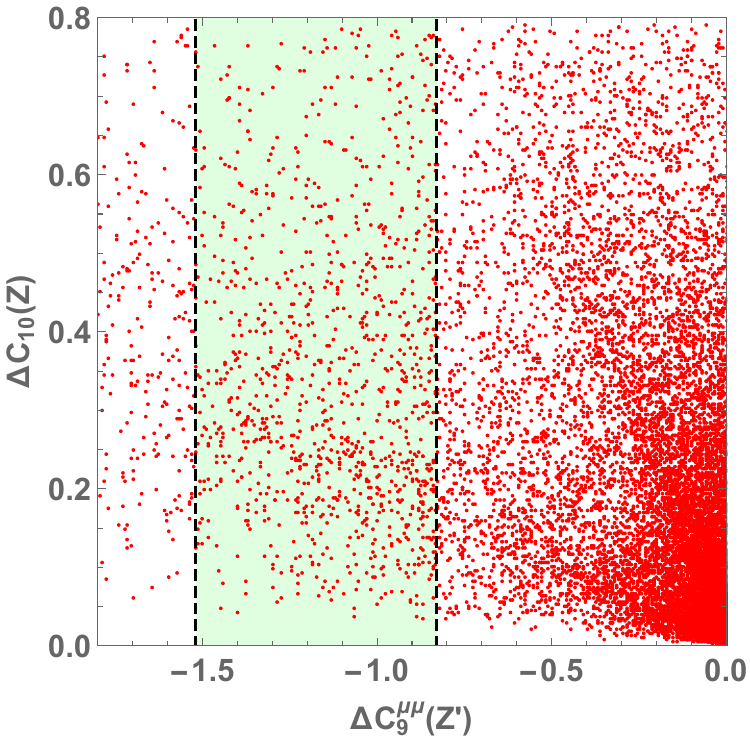} \quad
\includegraphics[width=60.0mm]{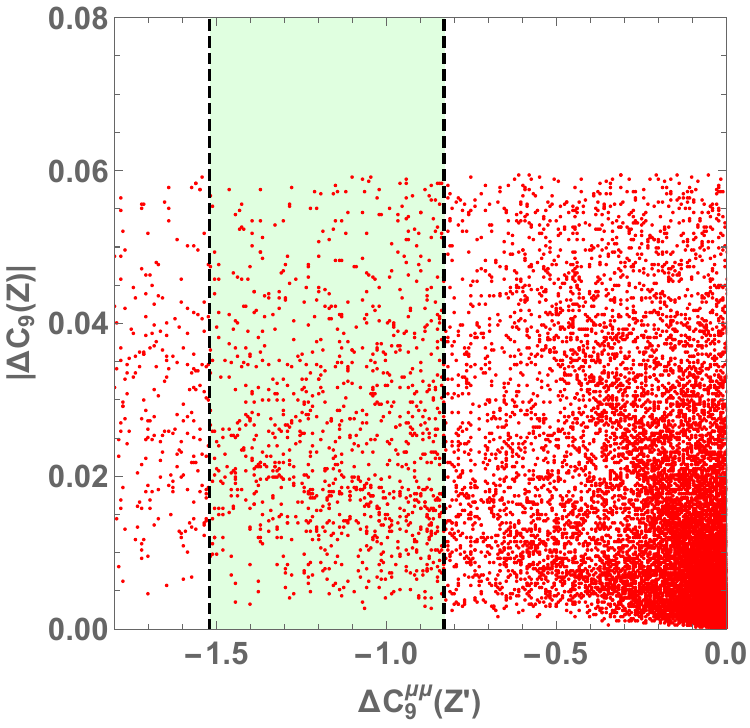}
\caption{Left: scattering plot on $\{ \Delta  C_9^{\mu \mu}(Z'), \Delta  C_{10}(Z) \}$ plain where each point satisfies phenomenological constraints except for $\Delta  C_{9,10}$. Right the same plot on $\{\Delta  C_9^{\mu \mu}(Z'), |\Delta  C_{9}(Z)| \}$ plain.} 
  \label{fig:C9C10}
\end{center}\end{figure}

\begin{figure}[t]
\begin{center}
\includegraphics[width=60.0mm]{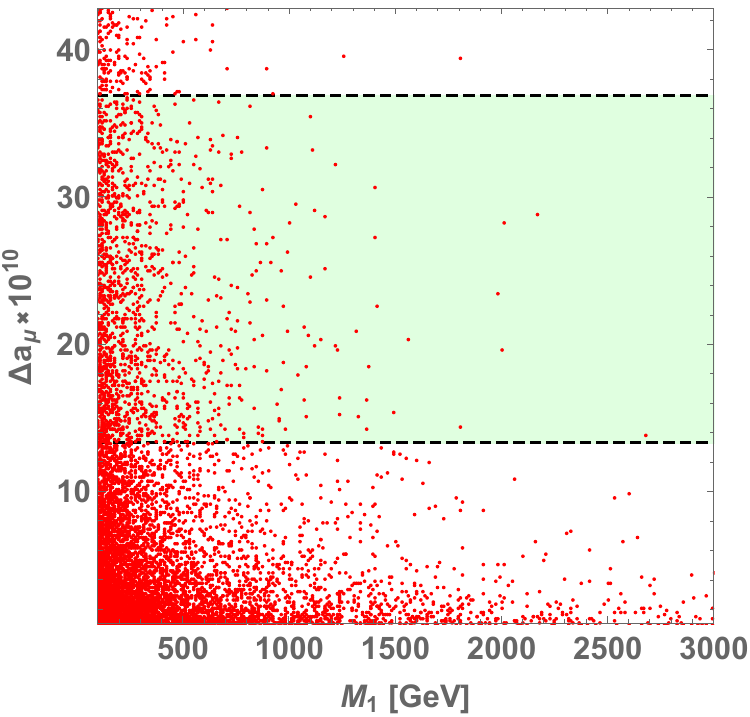}
\caption{$M_1$ versus $\Delta a_\mu$ for the parameter sets same as the previous plots.  } 
  \label{fig:amu}
\end{center}\end{figure}

In Fig.~\ref{fig:DM-Zp}, we show the allowed parameter region for DM mass $m_{\chi_\ell}$ 
and $m_{Z'}$.
The relic density is explained by $\chi \chi^* \to Z' \to \{\mu^+ \mu^-, \tau^+ \tau^-, \nu_{\mu,\tau} \bar{\nu}_{\mu,\tau}\}$ process for the points around $m_{Z'} \sim 2 m_\chi$
due to enhancement of the annihilation cross section. 
On the other hand $\chi \chi^* \to \mu^+ \mu^-$ via Yukawa interaction explains the relic density for the points which do not give resonant enhancement of the $Z'$ exchange process.

\begin{figure}[t]
\begin{center}
\includegraphics[width=60.0mm]{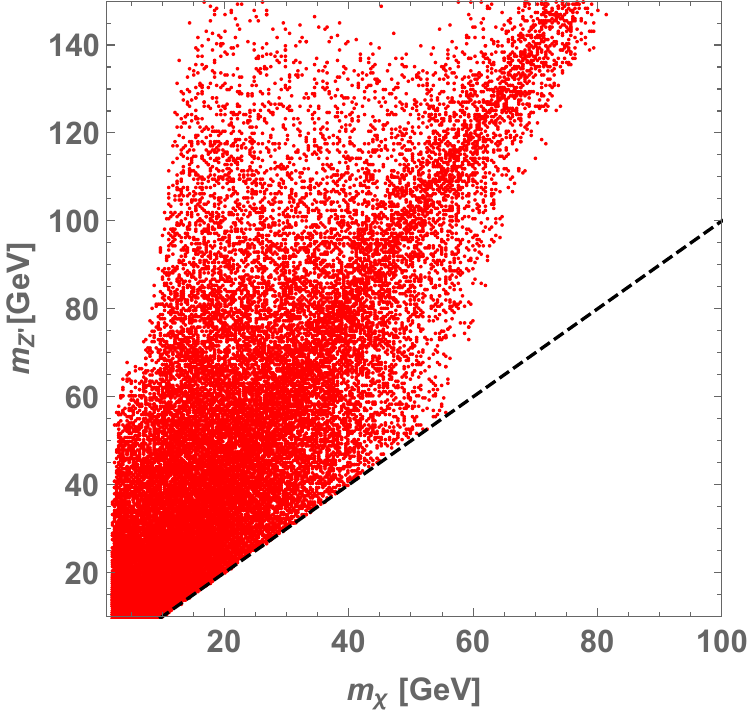}
\caption{$m_{\chi_\ell}$ versus $m_{Z'}$ in unit of GeV for the parameter sets same as the previous plots. 
} 
  \label{fig:DM-Zp}
\end{center}\end{figure}

\begin{figure}[t]
\begin{center}
\includegraphics[width=60.0mm]{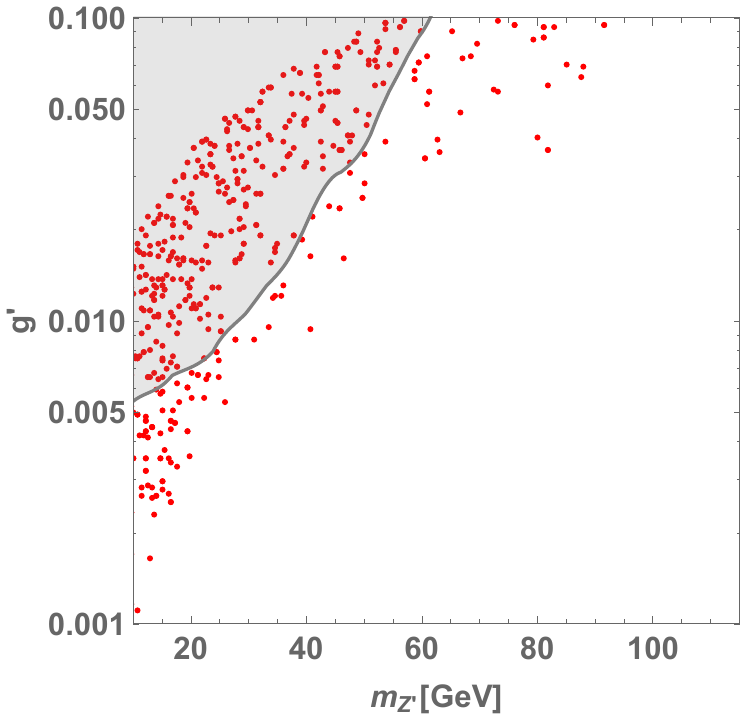}
\includegraphics[width=60.0mm]{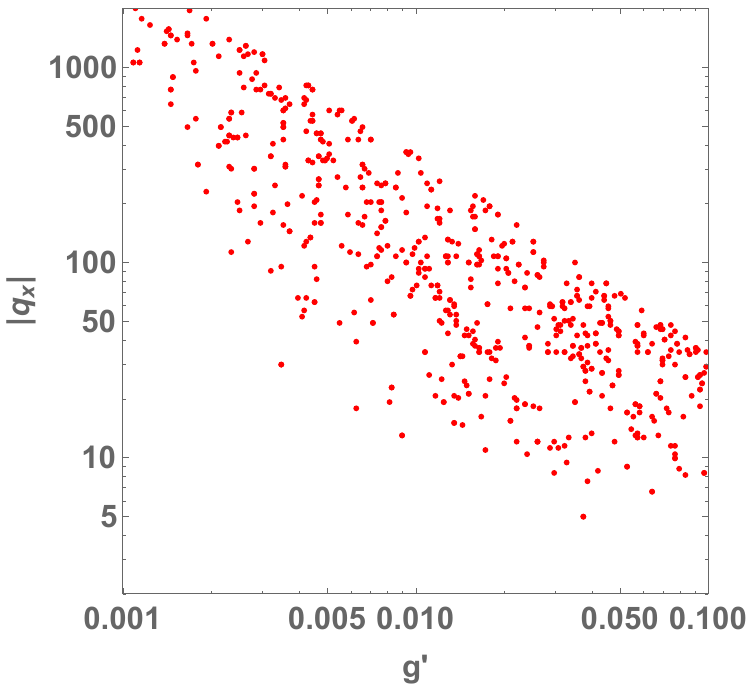}
\caption{ Left: Parameter points on $( m_{Z'}, g' )$ plane satisfying $0.83\le -\Delta C_9^{\mu\mu}\le 1.52$,  $\Delta a^{\rm new}_\mu = (25.1\pm 5.9)\times 10^{-10}$ and other phenomenological constraints. 
Shaded region is excluded by the LHC constraint from $pp \to \mu^+ \mu^- Z'(\to \mu^+ \mu^-)$ process~\cite{Sirunyan:2018nnz}. Right: the same points on $( g' , |q_x| )$ plane.} 
  \label{fig:gX}
\end{center}\end{figure}

Finally, in left plot of Fig.~\ref{fig:gX}, we show parameter points on $\{m_{Z'}, g_X\}$ plane satisfying $0.83\le -\Delta C_9^{\mu\mu}\le 1.52$,  $\Delta a^{\rm new}_\mu = (25.1\pm 5.9)\times 10^{-10}$ 
and other phenomenological constraints. In this region the most important one is the   
LHC constraint from $pp \to \mu^+ \mu^- Z'(\to \mu^+ \mu^-)$ process~\cite{Sirunyan:2018nnz}.
We find that $m_{Z'} \lesssim m_{Z}$ region is preferred to explain experimental anomalies where some parameter region is excluded by the LHC constraint for $m_{Z'} < 60$ GeV.
The gauge coupling should be small for small $m_{Z'}$ region to avoid the LHC constraint,
and large $|q_X|$ is required to explain $B$-anomalies in the region. 
In the right plot in Fig.~\ref{fig:gX} we show correlation between $g'$ and $|q_x|$ for the same parameter sets as the left plot;
both negative and positive values of $q_x$ appear as same amount and we only show absolute values.
We then find $g' |q_x| \sim \mathcal{O}(1)$ is required to obtain sizable $\Delta C_9^{\mu \mu}$.
Thus the region with $m_{Z'} > 60$ GeV would be more natural to explain $b \to s \mu^+ \mu^-$ anomalies since required $|q_x|$ is not too large.  
The remaining allowed region would be further tested in future at the LHC from the search for  
$pp \to \mu^+\mu^-Z'(\to \mu^+ \mu^- , \nu_{\mu,\tau} \bar \nu_{\mu,\tau} )$ and 
$pp \to \nu_{\mu,\tau} \bar \nu_{\mu,\tau} Z' (\to \mu \bar{\mu})$: namely 
4 muons or dimuon + missing transverse momentum.

\section{Summary and Conclusions}
In this paper, we have proposed a $U(1)_{\mu-\tau}$ gauged model 
to explain recent anomalies; the muon anomalous magnetic moment and $b\to s\mu^+ \mu^-$ 
at the same time, satisfying several constraints. Also, we have considered a complex scalar boson 
DM candidate ($\chi_l$) that mainly couples to leptons.
In our model $b\to s\mu^+ \mu^-$ anomalies can be explained by loop induced interactions among quarks and $Z'$.  We also estimated  one-loop diagrams associated with vector-like quark and $Z(\gamma)$ vertices that contribute to Wilson coefficients $C_{9(10)}$ and $C_7$ to check 
consistency of our model. 
We then find that $C_{9(10)}$ from $Z$ mediation and $C_7$ induced by vector-like quark loop can be small, and we can explain $b \to s \mu^+  \mu^-$ anomalies consistently. 
In our numerical analysis, we have found allowed regions to explain anomalies while satisfying phenomenological constraints. 
This model would be test soon in future measurement such as $pp \to \mu^+\mu^-Z'
(\to \mu^+ \mu^-)$ signal search at the LHC.
We also have the processes $pp \to \mu^+ \mu^- Z'(\to \nu_{\mu,\tau} \bar \nu_{\mu,\tau}  )$ 
and $pp \to \nu_{\mu,\tau} \bar \nu_{\mu,\tau} Z' (\to \mu^+ \mu^-)$ giving signal of 
$\mu^+ \mu^-$ with missing transverse momentum that is also searched for at the LHC.
In addition our model can be tested by searching for vector-like fermions at the LHC where vector-like quarks and leptons can be pair produced providing "dijet + missing transverse momentum" 
and  "dilepton + missing transverse momentum" respectively.

\section*{Acknowledgments}
The work of P.K. is supported in part by KIAS Individual Grants under Grant No.PG021403,  and by National Research Foundation of Korea (NRF) Research Grant NRF- 2019R1A2C3005009. 
The work of H.O. is supported by an appointment to the JRG Program at the APCTP through 
the Science and Technology Promotion Fund and Lottery Fund of the Korean Government, 
and also by the Korean Local Governments - Gyeongsangbuk-do Province and Pohang City 
H. O. is sincerely grateful for the KIAS members. 

\begin{appendix}

\section{Vector-like fermion contributions to electroweak vacuum polarization diagrams}
We calculate electroweak vacuum polarization diagrams with vector-like fermions and scalar bosons, 
and summarize analytic form of their contributions to estimate oblique parameters. 
We can write the vacuum polarizations for $Z$ and $W$ contributing to oblique parameters such that
\begin{align}
& \Pi_Z^{\mu \nu} = g^{\mu \nu} \frac{e^2}{c_W^2 s_W^2} (\Pi_{33}(q^2) - 2 s_W^2 \Pi_{3Q} (q^2) 
- s_W^4 \Pi_{QQ}(q^2)), \\
& \Pi^{\mu \nu}_W = g^{\mu \nu} \frac{e^2}{s_W^2} \Pi_\pm(q^2),
\end{align}
where $q$ is four momentum carried by gauge bosons.

Non-zero contributions to $\Pi_{33} (q^2)$ are summarized as follows;
\begin{align}
& \Pi^{E_1E_1}_{33}(q^2) = - \frac{s_c^4}{16 \pi^2} F(q^2, M_1^2,M_1^2), \\
& \Pi^{E_2E_2}_{33}(q^2) = - \frac{c_c^4}{16 \pi^2} F(q^2, M_2^2,M_2^2), \\
& \Pi^{E_1E_2}_{33}(q^2) = - \frac{s_c^2 c_c^2}{8 \pi^2} F(q^2, M_1^2,M_2^2), \\
& \Pi^{N'N'}_{33}(q^2) = - \frac{1}{16 \pi^2} F(q^2, M_{N'}^2,M_{N'}^2), \\
& \Pi^{U'U'}_{33}(q^2) = - \frac{1}{16 \pi^2} F(q^2, M_{U'}^2,M_{U'}^2), \\
& \Pi^{D'D'}_{33}(q^2) = - \frac{1}{16 \pi^2} F(q^2, M_{D'}^2,M_{D'}^2), 
\end{align}
where the superscripts in the left sides indicate particles inside vacuum polarization 
diagrams, and $F(q^2,m^2,m'^2)$ is the loop function given by
\begin{align}
F(q^2,m^2,m'^2) = &  \int_0^1 dx dy \delta(1-x-y) \left(\frac{1}{\epsilon_{\rm \overline{MS}}} - \ln \left(\frac{\Delta}{\mu^2} \right) \right) \nonumber \\
& \times (2x (1-x) q^2 - x m^2 - y m'^2 + m m'), \\
\Delta = & -x (1-x) q^2 + x m^2 + y m'^2, \quad  \frac{1}{\epsilon_{\rm \overline{MS}}} \equiv \frac{2}{\epsilon} - \gamma - \ln(4 \pi),
\end{align} 
where $\mu$ is auxiliary parameter having mass dimension. Dependence on $\mu$ is canceled when we calculate $S,T$-parameter from Eqs. (28) and (29) in Sec. III D.

Similarly we obtain non-zero contributions to $\Pi_{3Q, QQ, \pm}$:
\begin{align}
& \Pi^{E_1E_1}_{3Q}(q^2) = - \frac{s_c^2}{8 \pi^2} F(q^2, M_1^2,M_1^2), \\
& \Pi^{E_2E_2}_{3Q}(q^2) = - \frac{c_c^2}{8 \pi^2} F(q^2, M_2^2,M_2^2), \\
& \Pi^{U'U'}_{3Q}(q^2) = - \frac{1}{12 \pi^2} F(q^2, M_{U'}^2,M_{U'}^2), \\
& \Pi^{D'D'}_{3Q}(q^2) = - \frac{1}{24 \pi^2} F(q^2, M_{D'}^2,M_{D'}^2), 
\end{align}
\begin{align}
& \Pi^{E_1E_1}_{QQ}(q^2) = - \frac{1}{4 \pi^2} F(q^2, M_1^2,M_1^2), \\
& \Pi^{E_2E_2}_{QQ}(q^2) = - \frac{1}{4 \pi^2} F(q^2, M_2^2,M_2^2), \\
& \Pi^{U'U'}_{QQ}(q^2) = - \frac{1}{9 \pi^2} F(q^2, M_{U'}^2,M_{U'}^2), \\
& \Pi^{D'D'}_{QQ}(q^2) = - \frac{1}{36 \pi^2} F(q^2, M_{D'}^2,M_{D'}^2), 
\end{align}
\begin{align}
& \Pi^{E_1N'}_{\pm}(q^2) = - \frac{s_c^2}{8 \pi^2} F(q^2, M_1^2,M_{N'}^2), \\
& \Pi^{E_2N'}_{\pm}(q^2) = - \frac{c_c^2}{8 \pi^2} F(q^2, M_2^2,M_{N'}^2), \\
& \Pi^{U'D'}_{\pm}(q^2) = - \frac{1}{8 \pi^2} F(q^2, M_{U'}^2,M_{D'}^2). 
\end{align}

Finally non-zero contributions from new scalar particles 
beyond the SM are given by 
\begin{align}
& \Pi_{33}^{h,H} =- \frac{\sin^2 \alpha}{32 \pi^2} \int_0^1 dx \left[2 m_Z^2 \ln \left( \frac{\Delta_h}{\Delta_H} \right) + \Delta_h \left(\frac{1}{\epsilon_{\rm \overline{MS} }} - \ln \left(\frac{ \Delta_h}{\mu^2} \right) \right)- \Delta_H \left(\frac{1}{\epsilon_{\rm \overline{MS} }} - \left(\frac{ \Delta_H}{\mu^2} \right) \right) \right], \nonumber  \\
& \Pi_{\pm}^{h,H} =- \frac{\sin^2 \alpha}{32 \pi^2} \int_0^1 dx \left[2 m_W^2 \ln \left( \frac{\Delta_h}{\Delta'_H} \right) + \Delta'_h \left(\frac{1}{\epsilon_{\rm \overline{MS} }} - \left(\frac{ \Delta'_h}{\mu^2} \right) \right)
- \Delta'_H \left(\frac{1}{\epsilon_{\rm \overline{MS}}} - \left(\frac{ \Delta'_H}{\mu^2} \right) \right) \right], \nonumber \\ 
&\Delta_{h(H)} = x(1-x) q^2 + x m^2_{h(H)} + (1-x) m_Z^2,  \nonumber \\
& \Delta'_{h(H)} = x(1-x) q^2 + x m^2_{h(H)} + (1-x) m_W^2.
\end{align}  
This completes the new contributions to the vacuum polarization tensors of  electroweak gauge bosons in the $U(1)_{\mu - \tau}$ model considered in this paper. 
Then the $S$- and $T$-parameters can be obtained from Eqs. (28) and (29) in Sec. III D; 
divergent part proportional to $\epsilon_{\rm SM}$ will be canceled when we calculate oblique parameters by those equations.

\end{appendix}

\end{document}